\documentclass[%
 reprint,
 superscriptaddress,
 amsmath,amssymb,
 aps,
]{revtex4-2}

\usepackage{graphicx}
\usepackage{dcolumn}
\usepackage{bm}
\usepackage{amssymb}
\usepackage{siunitx}
\usepackage{upgreek}

\begin{document}

\preprint{APS/123-QED}

\title{Mapping Photocathode Quantum Efficiency with Ghost Imaging}

\author{K.~Kabra}
\email{krish97@g.ucla.edu}
\affiliation{Department of Physics and Astronomy, UCLA, Los Angeles, California 90095, USA}
\affiliation{Department of Electrical Engineering, UCLA, Los Angeles, California 90095, USA}

\author{S.~Li}
\email{siqili@slac.stanford.edu}
\affiliation{SLAC National Accelerator Laboratory, Menlo Park, California, 94025, USA}

\author{F.~Cropp}
\affiliation{Department of Physics and Astronomy, UCLA, Los Angeles, California 90095, USA}

\author{T.~J.~Lane}
\affiliation{SLAC National Accelerator Laboratory, Menlo Park, California, 94025, USA}

\author{P.~Musumeci}
\affiliation{Department of Physics and Astronomy, UCLA, Los Angeles, California 90095, USA}

\author{D.~Ratner}
\affiliation{SLAC National Accelerator Laboratory, Menlo Park, California, 94025, USA}

\date{\today}

\begin{abstract}
Measuring the quantum efficiency (QE) map of a photocathode injector typically requires laser scanning, an invasive operation that involves modifying the injector laser focus and rastering the focused laser spot across the photocathode surface. Raster scanning interrupts normal operation and takes considerable time to setup. In this paper, we demonstrate a novel method of measuring the QE map using a ghost imaging framework that correlates the injector laser spatial variation over time with the total charge yield. Ghost imaging enables passive, real-time monitoring of the QE map without manually modifying the injector laser or interrupting injector operation. We first demonstrate the method at the UCLA Pegasus photoinjector with the help of a digital micromirror device (DMD) and a piezoelectric mirror to increase our control of the overall transverse variance of the illumination profile. The reconstruction algorithm parameters are fine-tuned using simulations and the results are validated against the ground truth map acquired using the traditional rastering method. Finally, we apply the technique to data acquired parasitically from the LCLS photoinjector, showing the feasibility of this method to retrieve a QE map without interrupting normal operation. 

\end{abstract}

\maketitle

\section{Introduction}
Modern electron accelerators use photoemission to generate high brightness electron beams~\cite{fraser1986new,serafini1997envelope,dowell2016}. In this process, an optical drive laser strikes the surface of a photocathode, emitting electrons due to the photoelectric effect. The quantum efficiency (QE) map of a photocathode surface determines the spatial variation of the electron yield for a given incident photon flux.
The QE map of a photocathode is typically not uniform and deteriorates over time due to experimental conditions. For example, localized hot spots in the drive laser can burn the surface, which leads to degraded QE in a localized area.
Monitoring the QE map provides useful information for drive laser shaping to compensate for QE non-uniformities, which is crucial for obtaining the low emittances required for high-brightness applications \cite{Feng_BNL,Feng_LCLS}. Typically, the QE map is measured by focusing the drive laser to a small spot size and scanning it across the cathode surface.  In this configuration, the emitted charge at each location of the focused laser spot maps out the QE, assuming the emitted charge is below the space charge limit to preserve the linearity of the measurement.
This technique is by nature invasive to normal operation, as it changes the optical setup in the drive laser to focus the laser beam on the cathode surface. The resolution is limited by the focused laser spot size.

Another technique that has been successfully applied to measure the QE map~\cite{riddick2013photocathode} introduces a digital micromirror device (DMD) in the drive laser beam path. In this case, step size and spot size in the scans can be independently controlled with minimal optical realignment. Here, the resolution depends by the size of the micromirrors, the optical (de)magnification of the imaging system from the DMD to the cathode, and the signal-to-noise ratio of the charge measurement. Nevertheless, the use of the DMD (which has lower damage threshold than standard mirrors) significantly limits the amount of laser power that can be used for normal cathode illumination, and measurement is still invasive. 

In this work, we present a novel method to measure the QE map using classical ghost imaging.
Ghost imaging is an experimental technique that extracts spatial information from a single-pixel camera (also known as a ``bucket'' detector). 
In classical ghost imaging, an illumination source is typically split into two arms: one arm going to the sample under analysis and then to the bucket detector, the other arm reaching a pixelated detector. By correlating the bucket detector reading, i.e. the total emission from the sample, and the spatial variation of the illumination, one can reconstruct the spatial structure of the sample. Implicit in this is the assumption that the spatial profile of the incident illumination varies shot-to-shot so that measurements are sufficiently independent. Classical ghost imaging in the spatial domain has been demonstrated experimentally with various sources of illumination, including visible light, optical lasers, x-rays, atoms, and electrons~\cite{pelliccia2016,yu2016,khakimov2016ghost,li2018electron}.

Ghost imaging offers several advantages over the state-of-the-art raster scan. For example, simultaneously illuminating multiple pixels (i.e. ``multiplexing'') improves the signal-to-noise ratio when the dominant source of uncertainty is an overall detector noise independent from the number of pixels in the image (the so-called Fellgett's advantage \cite{Fellgett}). Furthermore, multiplexing also enables the use of compressive sensing \cite{candes,katz2009compressiveGhost,wallerMultiplexedFourierPtychography}, which can reduce the number of measurements needed to reconstruct a target for sparse samples. Finally, and crucially for QE mapping, the ghost imaging framework can use the intrinsic noise in the illumination pattern of the photocathode drive laser to implement a passive measurement method that avoids disturbing normal operation. See \cite{lane2019advantages} for a summary of the advantages of multiplexing.

It is this final advantage, the ability to measure machine parameters passively, that makes the application of ghost imaging to QE mapping particularly attractive for user facilities. In this paper we first demonstrate the idea with a proof-of-principle measurement on a dedicated test facility, and then show results from a passive QE measurement acquired during beam delivery at the Linac Coherent Light Source (LCLS).

\section{Classical Ghost Imaging for Cathodes}

Ghost imaging requires a varying, known illumination pattern paired with synchronous measurements of a bucket detector. In typical ghost imaging experiments, the variation is generated by inserting a random pattern in the beam path (see e.g. \cite{yu2016}) or by directly controlling the illumination pattern \cite{shapiro08}.  For the latter case, one can introduce variation by placing a spatial light modulator, such as a DMD, into the beam path of the incident light. Imposing user-programmed masks on the incident light eliminates the need for an arm to measure the spatial profile of the incident light, and hence is called ``computational ghost imaging.''



The ghost imaging reconstruction algorithm can be boiled down to solving a linear matrix inversion problem:
\begin{equation}
    \mathbf{b}=\mathbf{A}\mathbf{x},
\label{eq:b=Ax}
\end{equation}
where $\mathbf{b}$ is the bucket detector reading, 
$\mathbf{A}$ is the matrix of the incident light spatial profiles, and $\mathbf{x}$ is the unknown sample.
If we denote the number of measurements by $m$ and number of pixels by $p$, then we have $\mathbf{b} \in \mathbb{R}^{m\times1}$, $\mathbf{A} \in \mathbb{R}^{m\times p}$, and $\mathbf{x} \in \mathbb{R}^{p\times 1}$.
In most ghost imaging problems,
one can choose from a collection of established algorithms to obtain an approximate solution to the unknown $\mathbf{x}$ in a linear system of the form of Eq.~\ref{eq:b=Ax}.

Because the total emitted charge from a photocathode is proportional to the product of the drive laser spatial profile and the QE map, measuring the QE fits the scheme of classical ghost imaging.
Here, the bucket detector reading is the total charge emission from the cathode, and the sample to be imaged is the QE map.
By relating the set of varying drive laser spatial profiles ($\mathbf{A}$) to its respective total charge emission measurements ($\mathbf{b}$), it is possible to retrieve the spatial features of the QE map ($\mathbf{x}$).
A practical implementation of QE ghost imaging requires only synchronized measurements of the drive laser profile and the emitted charge, which are already available at most accelerator facilities.

For Eq.~\ref{eq:b=Ax} to be solvable, the rows of $\mathbf{A}$ should be independent, i.e. the laser profile should change from shot-to-shot. In computational ghost imaging, the variation is provided by intentionally controlling the laser profile, e.g. with a DMD \cite{pietsunDMDSpeckleIllum}. However, ghost imaging can also exploit natural variation in the drive laser profile to measure a QE map without interrupting normal accelerator operation. Even a small degree of variation, if uncorrelated, can be utilized by acquiring a larger number of examples. As data acquisition is completely passive, the possibility exists to collect large data sets parasitically over an extended period of time, capturing more variation from the natural spatial ``jitter'' of the drive laser. 


\section{Experimental Demonstration}

Though our goal is to exploit the random variation of the cathode laser, we start by intentionally varying the laser illumination to evaluate the limits of the reconstruction and the required variation in the illumination pattern.

At the UCLA Pegasus beamline, we insert a DMD into the laser path to control digitally the drive laser profile. The use of dynamical and digitally controllable laser shaping techniques for the drive laser has been demonstrated in photoinjectors, as in Refs.~\cite{jared,li2017ultraviolet,li2018electron}. In our case, the DMD enables quantitative comparison between the traditional scanning method and our proposed ghost imaging method. First, we obtain a ground truth QE using a raster scan. Second, we generate a sequence of random masks on the drive laser profile.   
Finally, we use a mirror controlled by a piezoelectric motor located after the DMD to shift the position of the illumination pattern on the cathode, emulating the transverse position jitter in the laser. Artificially increasing the amount of jitter enables a test of the technique using fewer measurements. 
Post-experiment, we analyze the laser variation for each data set, and compare QE map reconstructions between the traditional raster scan and the two variations that utilize the ghost imaging method. We perform simulations using the experimental illumination patterns to guide the choice of the reconstruction algorithm hyper-parameters, as well as analyze the ability of random mask and laser jitter scans to retrieve QE maps. 

In the following section, we demonstrate a practical experimental implementation of the technique at the linac coherent light source (LCLS) where data is instead acquired in parasitic mode, using the intrinsic jitter of the drive laser profile to reconstruct the QE map of the photocathode.  

\subsection{Pegasus: Experimental Setup}

\begin{figure}
    \centering
    \includegraphics[width=\linewidth]{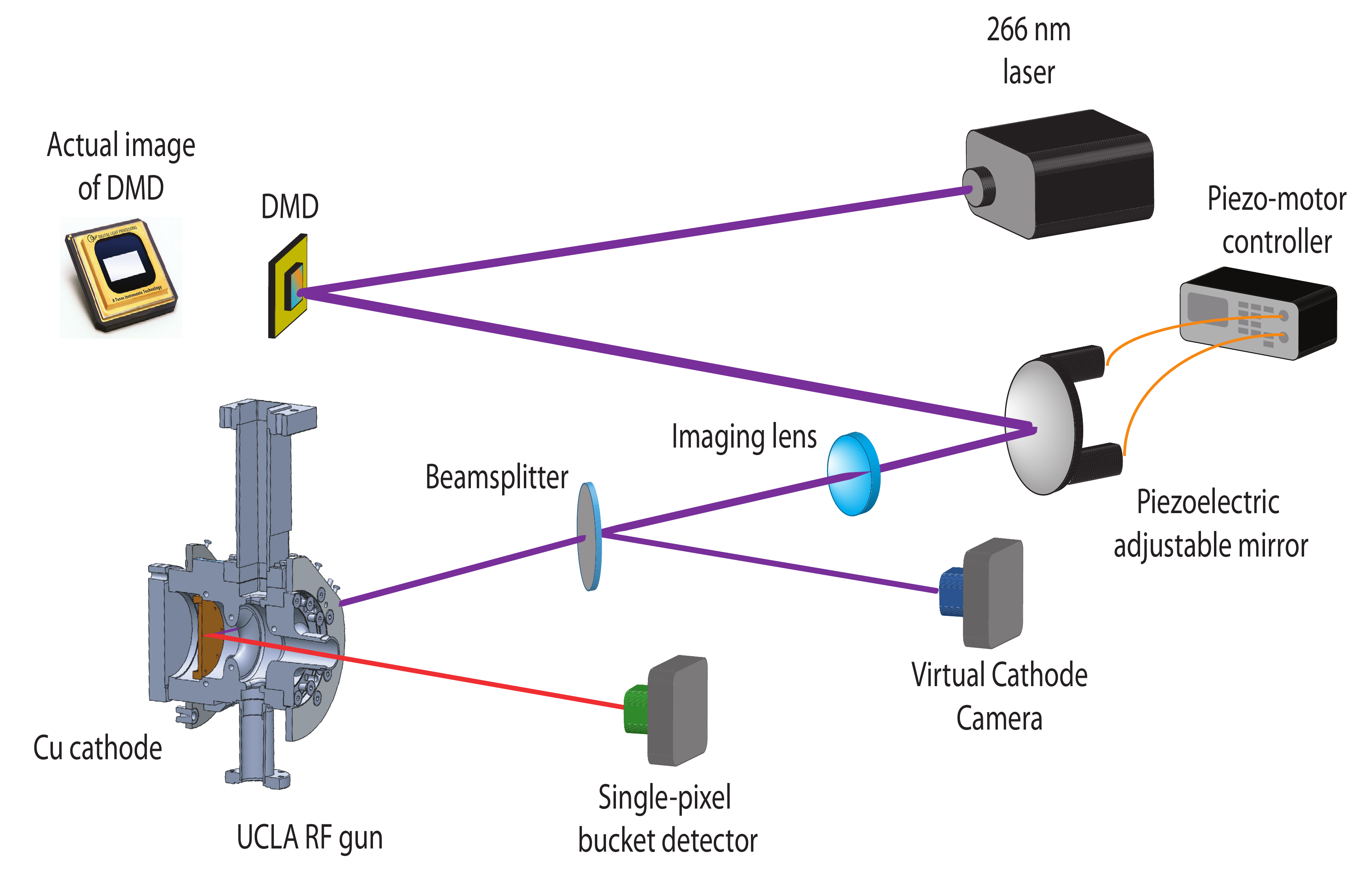}
    \caption{Simplified schematic of the UCLA experiment. A 266 nm laser illuminates 240 $\times$ 240 pixels on the DMD, which contains a programmable mask. The transversely-shaped laser then reflects off of a UV-enhanced metal mirror that is mounted on a piezoelectric kinematic mirror mount that can be electronically adjusted along two axes. The laser is then imaged onto a Cu cathode with a magnification factor of 3.8 using a lens with a 75 cm focal length. A small portion of the beam is split using a beam-splitter upstream of the cathode and is directed onto a screen located at the imaging plane of the lens and monitored by a virtual cathode camera (VCC).}
    \label{fig:UCLA_exp}
\end{figure}

The experiments at the UCLA Pegasus beam line \cite{alesini} use a 100 fs rms 266 nm laser pulse to strike a copper cathode in the high field of an S-band 1.6 cell radiofrequency (RF) gun to generate the 3.21 MeV kinetic energy electron beam. The transverse profile of the injector laser is controlled using a Texas Instruments' DLP-7000 DMD with a modified window to allow ultraviolet (UV) transmission, as discussed in \cite{li2017ultraviolet}. A 3 mm full width laser spot fully illuminates 240 x 240 pixels on the device. Due to the grating-like structure of the DMD, the shaped laser is dispersed into several diffraction orders and a pulse-front tilt (PFT) is introduced by the device. We use a 1800 lines/mm diffraction grating upstream of the DMD to compensate for this PFT. The brightest central order reflected from the DMD is selected and imaged onto the cathode with a magnification factor of 3.8 using a 75 cm focal length lens. A beam splitter upstream of the cathode directs a small portion of the laser to a screen placed at the imaging plane and is monitored by a standard CCD referred to as the virtual cathode camera (VCC). Because of the significant losses on the grating, DMD, and transport line, less than 10\% of the input UV energy reaches the cathode. The input UV laser is also limited to $<20~\upmu$J to avoid DMD damage.

After the RF gun, the electron beam is focused by a solenoid before a fluorescent screen that is imaged with a standard CCD camera. The electron beam and photoinjector laser profile images are collected synchronously using an external trigger and labeled using a digital timestamp. In post analysis, all background from the electron beam image is removed and a single-pixel ``bucket sum'' is obtained by integrating the signal. This bucket sum is directly proportional to the emitted charge.

The ground truth QE map is obtained using a raster scan data set. A single 16 $\times$ 16 DMD macropixel, defined as squares with widths of 16 pixels, was turned on for each scanning point. The corresponding emitted charge is obtained by integrating the electron image after background subtraction. Similarly, the total laser power for each beam shot is measured by integrating the intensity on the VCC image. The ground truth QE is mapped out by filling in the ratio of charge emission divided by the total laser power for each scanning data point on the VCC screen, as shown in Fig.~\ref{fig:UCLA_recon} (left).
The resolution associated with the size of the macropixel at the cathode plane corresponds to $57 \pm 5~\upmu$m, which is comparable to the resolution obtained in traditional scanning methods and is sufficient to capture the QE features on the photocathode used for this experiment.  

To test the ghost imaging technique, two main data sets were collected during the experiment. First, for the ``random'' data set, we displayed 350 random (binary) masks on the DMD. In each random mask, each of the 16 $\times$ 16 macropixels in the illuminated 240 $\times$ 240 pixel region was turned on with 50\% probability. Second, the ``jitter'' data set consisted of 361 measurements with all pixels on the DMD turned on. We varied the injector laser profile on the cathode by remotely steering the beam using a piezoelectric motor controlled mirror. The beam was steered by a maximum of 0.2 mm (nearly 25 $\%$ of the laser spot size) horizontally and vertically on the cathode from the starting reference position. Although this variation is larger than that observed in typical photoinjector setups, the number of measurements is orders of magnitude smaller than the data that can be collected passively at facilities such as LCLS.

For both data sets, the corresponding VCC images are cropped to maximally encompass the laser beam and then subsequently downsized to a 30$\times$30 grid. 
We downsize the VCC images to reduce the number of free parameters for our regression algorithm and avoid an underdetermined regression problem that makes solving Eq.~\ref{eq:b=Ax} more difficult. Downsizing can be avoided at the cost of taking more measurements, use of stronger regularization, and longer computational reconstruction times.  

\subsection{Pegasus: Data Analysis and Simulations}
\label{sec:UCLA_analysis}

In order to quantify the variation of the drive laser for a particular data set we consider two methods of analysis. The first method examines the coefficient of variation (CV), also known as the relative standard deviation, for each pixel across the VCC data set. The second method uses principal component analysis (PCA), a statistical procedure where the variables of a data set are transformed into an ordered orthogonal basis \cite{PCA_jolliffe_2011}. The first principal component (PC) correponds to an eigenvector of the data set containing the largest variance. PCA allows us to capture information concerning the linear independence of laser profiles across different measurements. If there are enough eigenvectors accounting for a majority of the total drive laser spatial variation, then there is a significant amount of laser variation that ensures shot-to-shot measurements are sufficiently independent.

\begin{figure}[t]
    \centering
    \includegraphics[width = \linewidth]{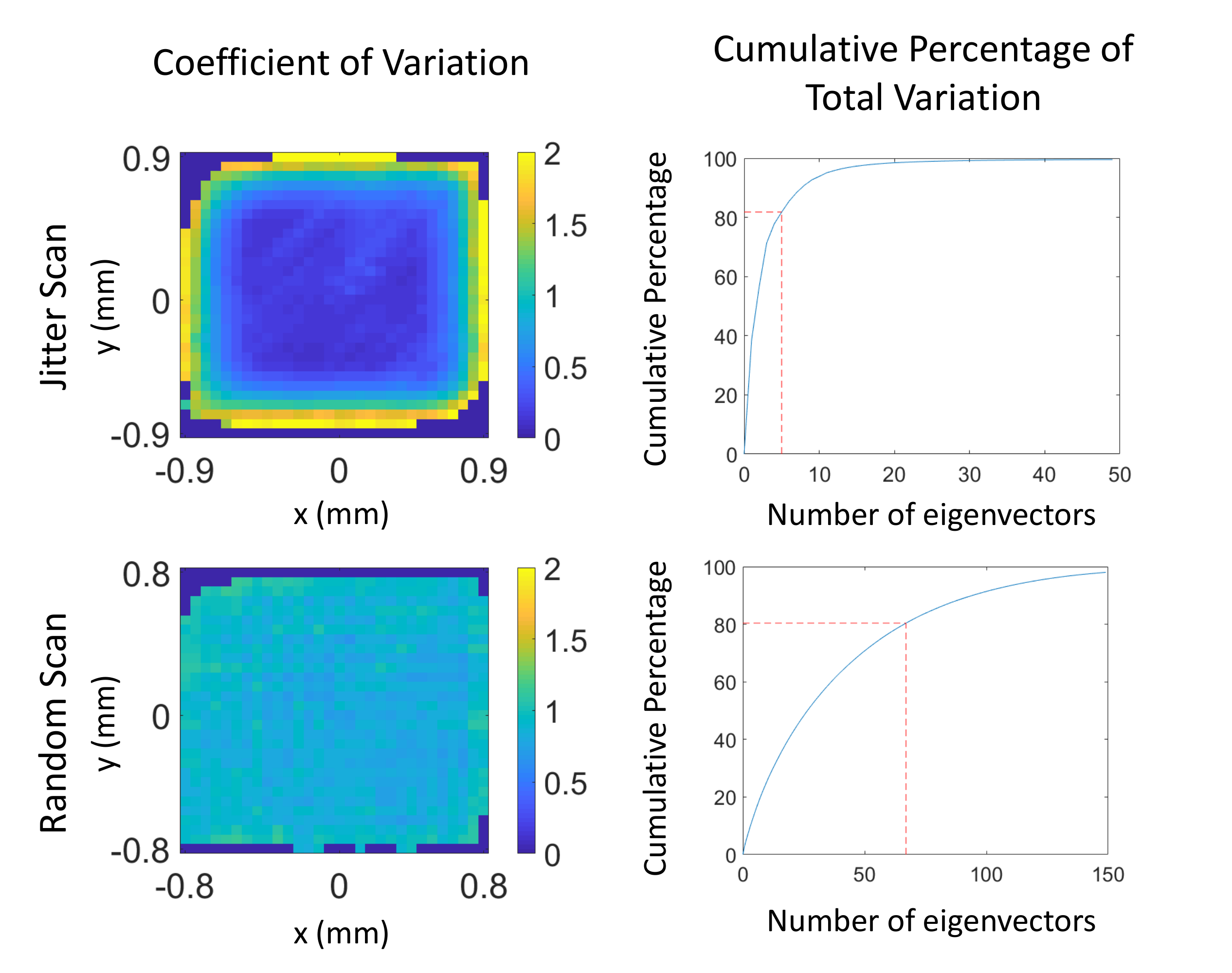}
    \caption{Drive laser variation analysis for UCLA jitter scans (top row) and random scans (bottom row). The left-column shows the CV for each pixel. Note that edge pixels that have a mean of approximately zero have been discarded to avoid division by zero errors. The right-column shows the cumulative percentage of total variation contained within the eigenvectors of the data set. The dashed lines indicates the number of eigenvectors containing 80\% of the total variation. This was 5 eigenvectors for the jitter scans and 67 eigenvectors for the random scans. The total number of eigenvectors for both data sets was 900.}
    \label{fig:VCCanalysisFig}
\end{figure}

The analysis for the UCLA random and jitter scans can be seen in Fig. \ref{fig:VCCanalysisFig}. As expected, the CV analysis shows that the random scan contains more laser variation across the region of interest whereas the jitter scan contains variation concentrated along the edges. We note that the checkerboard pattern in the random scan CV figure is an artifact of DMD damage. Moreover, the PCA results indicate that it takes more eigenvectors in the random scan to describe the same level of cumulative percentage of total variation than the jitter scan.

\begin{figure}[t]
    \centering
    \includegraphics[width = \linewidth]{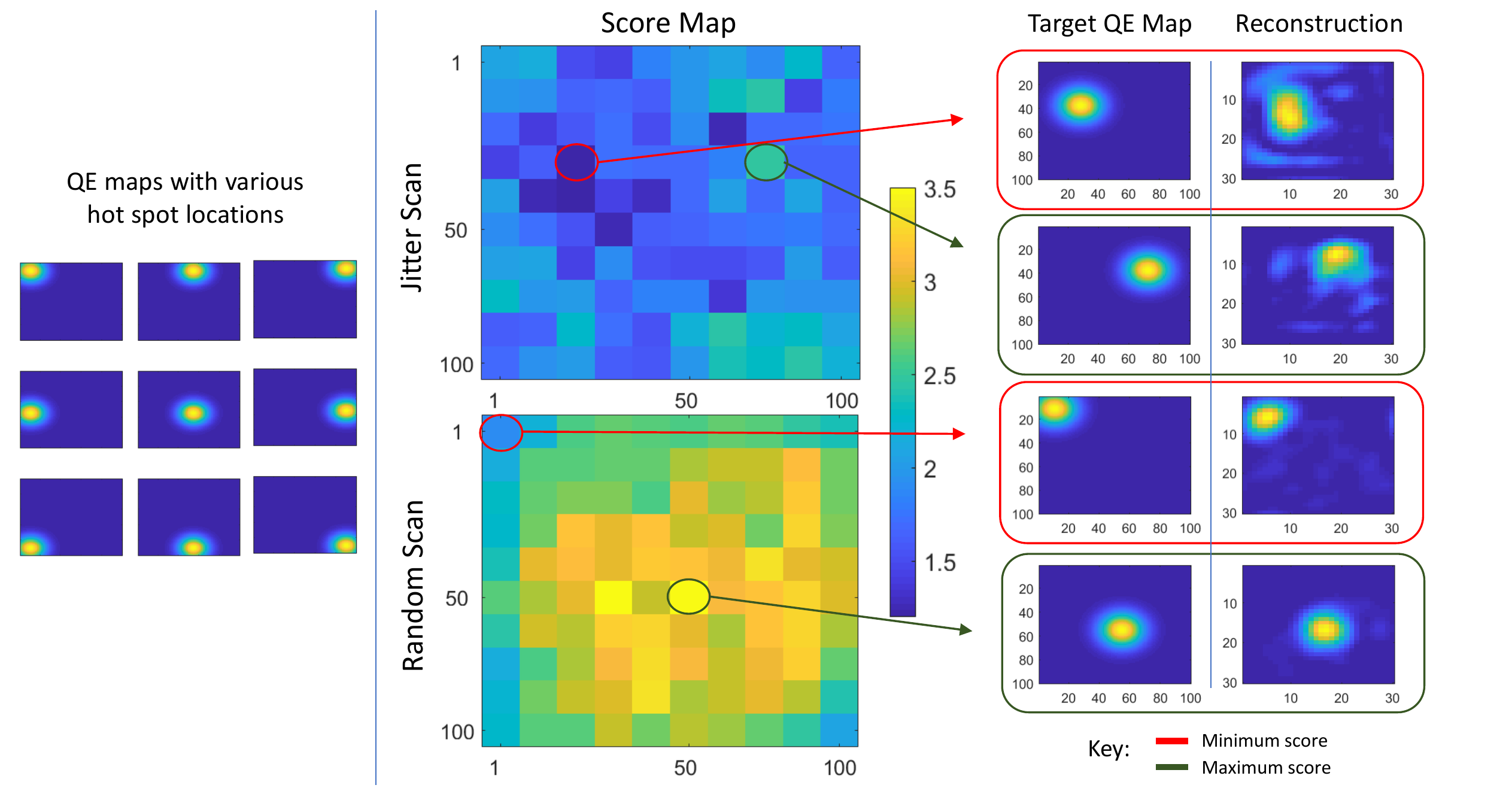}
    \caption{Simulation results using UCLA jitter (top) and random scan (bottom) data sets for a collection of QE maps each containing one 2D Gaussian-shaped hot spot with a centroid placed at a unique location. The left-most image shows a subset of the generated QE maps. The score maps (middle) represent the performance of a reconstruction with a target hot spot centered at the coordinate within the map. The score is defined in Eq. \ref{eqn:score}. The right-most figures show the best and worst scored reconstructions side-by-side with the target QE map. Note that the reconstructions have the same size of as the VCC images, and are upsized to match the dimensions of the target QE map during score calculations. Some of the jitter reconstructions includee artifacts, but both methods identify the QE hotspot. }
    \label{fig:simulationData}
\end{figure}

To clarify the effects of this difference in the illumination patterns, we conduct simulations using the UCLA jitter and random scan VCC data and multiple user-generated QE maps each containing one 2D Gaussian-shaped hot spot with a centroid placed at a unique location. The simulation method is discussed in Appendix~\ref{app:simulation}. The results of the simulation can be seen in Figure \ref{fig:simulationData}. A score map is used to visualize the performance of reconstructions for hot spots centered at a coordinate within the map. The score is defined to be: 
\begin{equation}
    S = \log{\frac{1}{MSE}}
    \label{eqn:score}
\end{equation}
where $S$ is the score and $MSE$ is the mean-square error between the target QE map and the reconstruction upsized to have the same dimensions as the target. With this definition, a higher score implies a more accurate reconstruction. The reconstructions for both the jitter and random scans appear to be well-behaved for QE hot spots centered in any arbitrary location. As expected the random mask data set provides more accurate reconstructions as it contains more spatial variation in the illumination series (see Fig. \ref{fig:VCCanalysisFig}). The jitter data set results in non-physical artifacts, but still recovers the cathode hot spot in each case. As we compare different experiments, it is important to note that the amount of variation required also greatly depends on the number of samples as well as the SNR in the bucket sum acquisition.

\begin{figure}[t!]
    \centering
    \includegraphics[width=\linewidth]{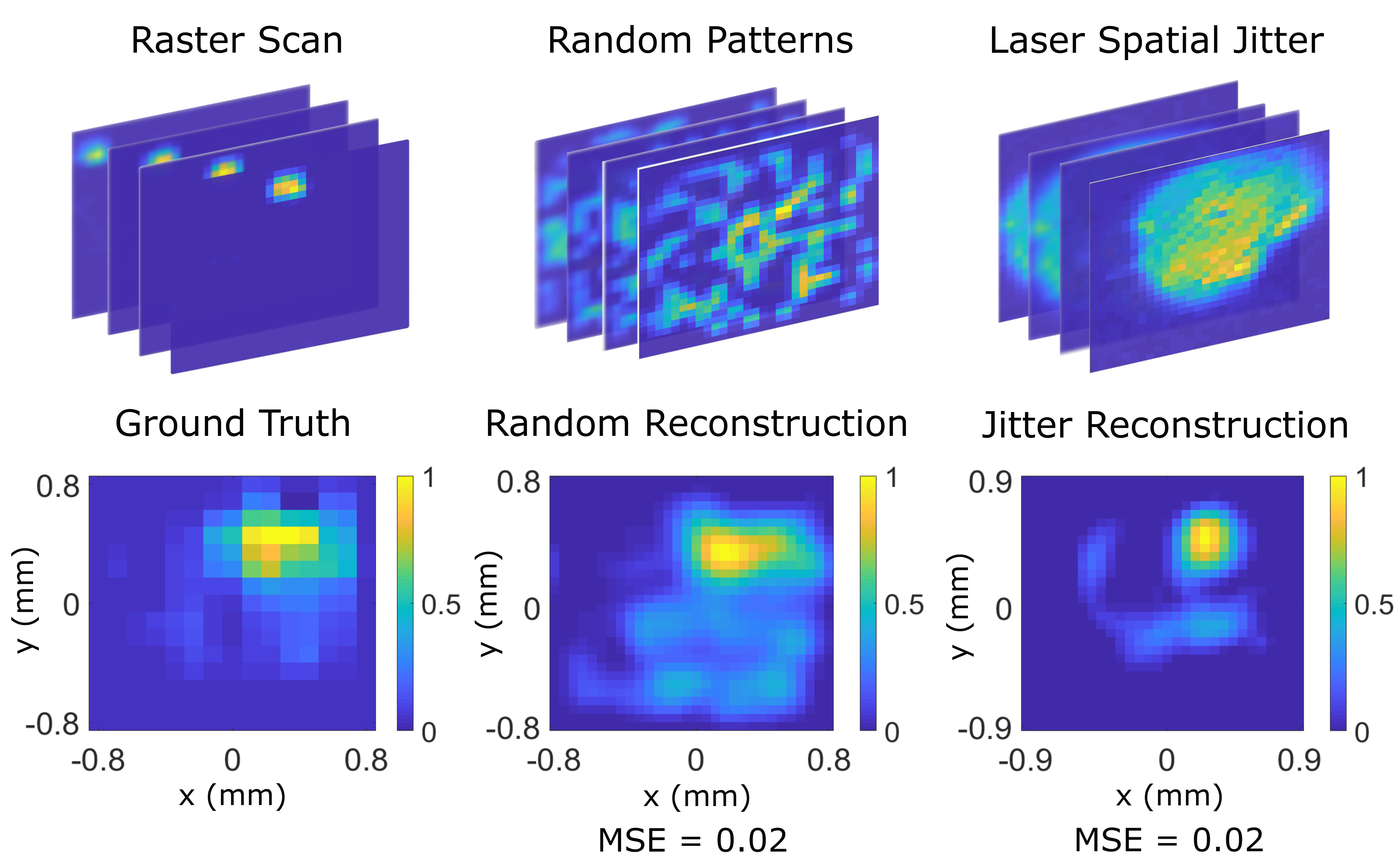}
    \caption{QE map reconstructions. The top row shows example laser profiles used to generate the QE maps seen on the bottom row. Left: Ground truth QE map obtained by raster scan. Middle: reconstruction using the random mask data set. Right: reconstruction using the jitter data set. The random and jitter data sets successfully reconstruct the hot spot seen in the upper right corner of the ground truth.}
    \label{fig:UCLA_recon}
\end{figure}

Finally, we reconstruct the QE map using the alternating direction method of multipliers (ADMM) algorithm \cite{boyd2011distributed}. The hyperparameters for ADMM are selected to optimize a simulated QE map consisting of a single Gaussian-shaped hot spot centered at the origin. The choice to use this map for the simulation comes from the fact that many photoinjector based accelerator facilities utilize a photocathode with a QE that contains a single hot spot (see for example \cite{cultreraCornellPhotocathode,bardayPhotocathode,stephanPITZ}). The reconstructions are presented in Fig.~\ref{fig:UCLA_recon}. 

The quality of the reconstructions is quantified by the MSE with the raster scan ground truth. Both the random and jitter data sets capture the major feature of a hot spot on the upper right corner, consistent with the ground truth. Moreover, the random data set captures low-level QE variations below the hot spot that are not captured by the raster scan. This is also partially captured by the jitter scan. This feature may be real and only resolvable by the random and jitter scans due to Fellgett's advantage. 

In sum, we are able to reconstruct the quantum efficiency of the cathode using either the DMD to create random patterns or passively using the jitter from the laser pointing. Using simulation, we show how the CV and PCA can help assess the amount of laser spatial variation needed for a reconstruction. As expected, the jitter data set performs worse than the random set, but is still capable of recovering major features in the QE map. In the next section we will show how the method translates to operation at LCLS, where the natural spatial jitter in the illumination is smaller, but it is possible to record orders of magnitude more data.

\subsection{LCLS: Parasitic Demonstration}




\begin{figure}[b!]
\centering  
    \includegraphics[width=\linewidth]{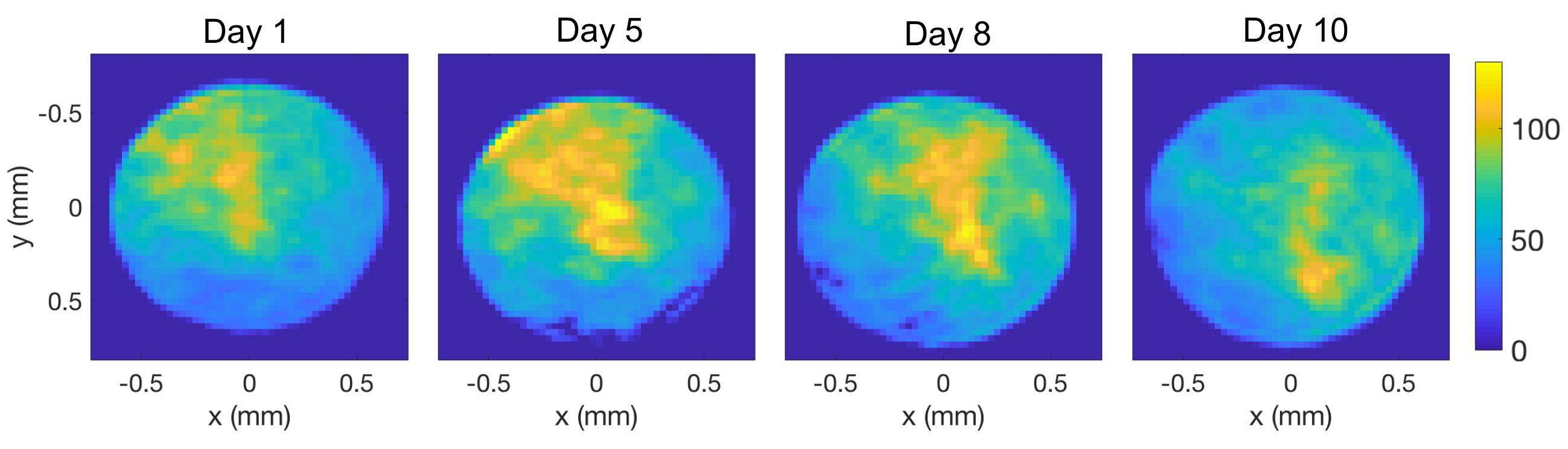}
    \caption{Examples of single shot LCLS injector laser profiles over the course of 10 days. Camera intensity is in arbitrary units.}
    \label{fig:LCLSlasers}
\end{figure}

At LCLS, the injector laser consists of a Ti:Sapphire laser system, producing 2~ps laser pulses at 760~nm wavelength. The infrared laser is then tripled to ultraviolet wavelength (253~nm) and generates electrons by striking a copper photocathode~\cite{akre2008commissioning}.
In normal operation, the laser spot size has a diameter of $\sim$1~mm on the photocathode, emitting electron bunches of~$\sim$250~pC.
Similarly to the UCLA Pegasus setup, before the injector laser reaches the cathode, it goes through a beam splitter where a small portion of the beam is imaged onto a VCC which captures the injector laser profile on the photocathode.
We record the injector laser profile synchronously with the corresponding charge measured by a beam position monitor.
Over the course of ten days, we took data parasitically for about two hours each day during both normal operation for XFEL and accelerator beam studies.
After filtering out shots where the charge drops below 100~pC or when the laser is shuttered, we obtain on average 6949 shots per day.
We crop the VCC images to encompass the full laser beam and downsize the images to a 20$\times$20 grid to reduce the number of free parameters.
In the downsized image, each pixel size corresponds to~$\sim$76~$\upmu$m, which is less than 10\% of the beam diameter and therefore retains the features in the laser beam.

\begin{figure}[b!]
\centering  
    \includegraphics[width=0.9\linewidth]{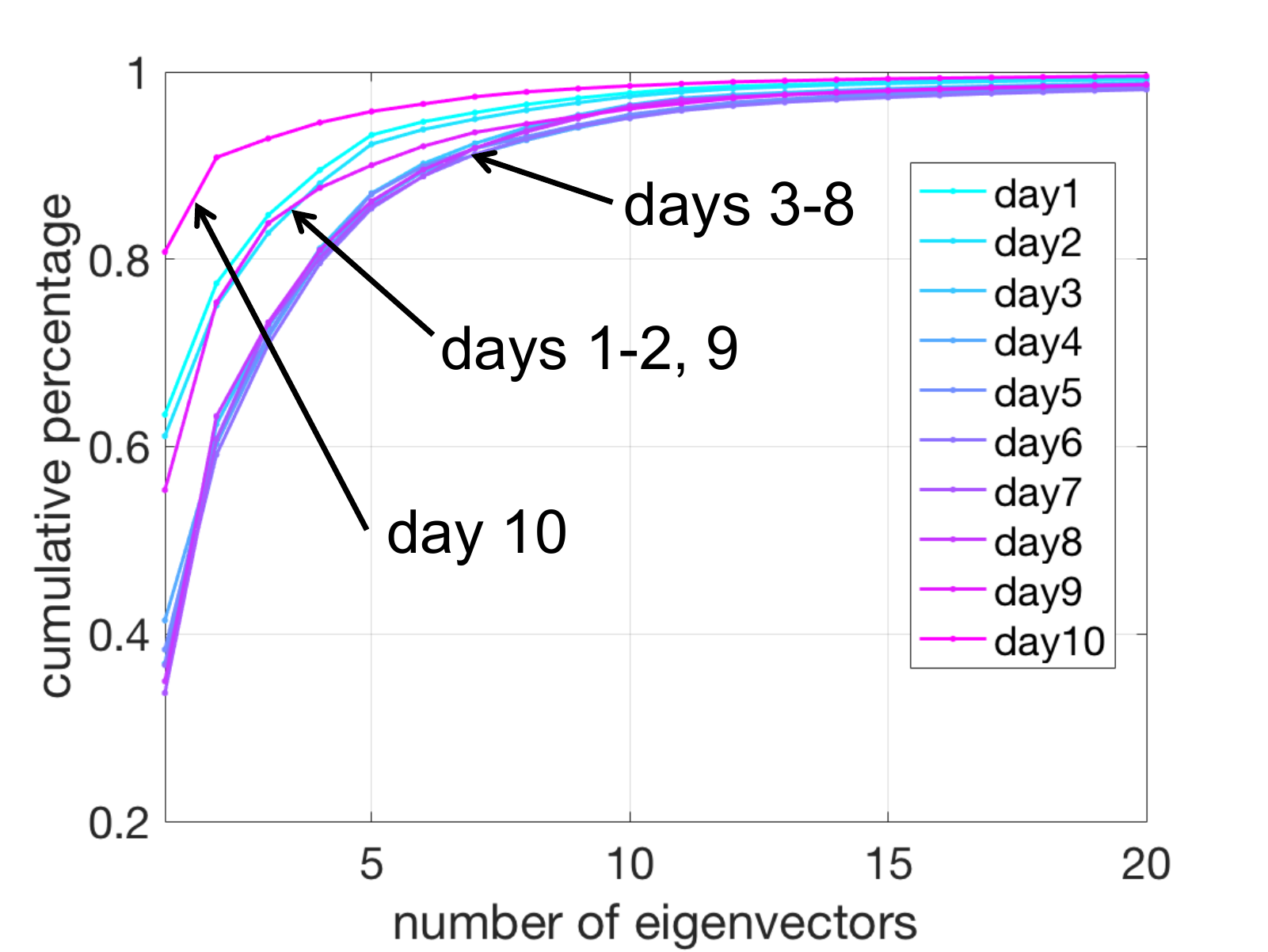}
    \caption{Cumulative variation explained by number of eigenvectors from PCA of the LCLS laser profiles. To capture 80\% of the total variation, it takes 1 to 5 eigenvectors depending on the day. The total number of eigenvectors is 400.}
    \label{fig:LCLS_pca}
\end{figure}

One challenge for collecting data parasitically is the charge-feedback system, which continuously tunes the injector laser power to maintain an average charge level around 250~pC. Because we do not have measurements of the underlying cause, we account for the charge feedback by normalizing the VCC images by the integrated laser intensity of a running average over 100 shots. Note that the normalization implicitly assumes that the drift in charge is not caused by changes to the QE map (which we assume to be fixed during the measurement) or the laser profile jitter (which is fast compared to the feedback). We have also verified that choosing to average over 100 shots does not significantly change the PCA curve as shown in Fig.~\ref{fig:LCLS_pca}, and, therefore, does not remove the shot-to-shot fluctuations in the laser profiles.

Figure~\ref{fig:LCLSlasers} shows the natural spatial jitter apparent in the LCLS injector laser on four of the ten days.
In order to assess the variation of the laser profiles, we apply the PCA decomposition method described in the Sec.~\ref{sec:UCLA_analysis} to the LCLS data.
Figure~\ref{fig:LCLS_pca} shows the PCA results of the LCLS data by day. 
It takes similar number of eigenvectors to cover 80\% of the total cumulative variation compared to the UCLA jitter scan, and significantly fewer eigenvectors compared to the UCLA random scan. This indicates that the natural jitter provides comparable variation as inducing jitter by steering mirrors and less variation than introducing random binary patterns. As shown later, this amount of variation is sufficient to produce a consistent QE reconstruction over days.
The PCA metric provides qualitatively the level of validity of the reconstructions using the data from different days.
Furthermore, from the steering mirror movement at LCLS we see there was large movement of the mirror between day 6 and 7 due to the accelerator's operational interruptions, which is known to cause shifts due to backlash in the steering motors. Although there has been no intentional movement of the laser position on the cathode, we estimate from the drift in the mirror motor movement (see Fig.~\ref{fig:LCLS_M2}) that the beam moved by approximately 10\% of the beam size.

Next, we conduct a simulation study with a user-generated QE map similar to that described in Appendix ~\ref{app:simulation}. We simulate the charge measurement by Eq.~\ref{eq:b=Ax_sim} using the measured VCC images and a Gaussian noise with an
\begin{figure}[h]\setlength{\hfuzz}{1.1\columnwidth}
\begin{minipage}[t]{0.99\textwidth}
\centering  
    \resizebox{0.8\textwidth}{!}{\includegraphics{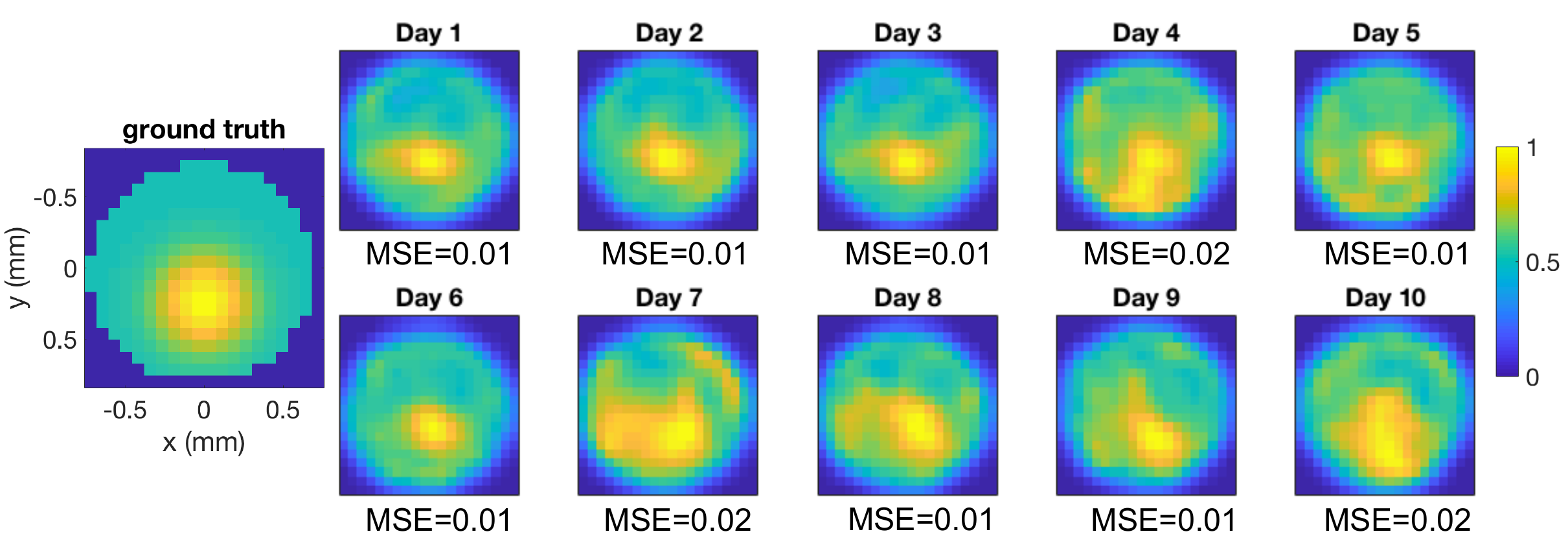}}
    \caption{Simulated ground truth (at left) and day-by-day reconstruction using recorded LCLS VCC images and simulated bucket measurements. A faint circular ring of high QE appears consistently on the aperture of the laser profile. This is in error with the simulated ground truth, implying that it is an artifact of the reconstruction algorithm.}
    \label{fig:LCLS_sim_day}
\end{minipage}
\end{figure}
\vspace{-15pt}
\begin{figure}[h!]\setlength{\hfuzz}{1.1\columnwidth}
\begin{minipage}[t]{0.99\textwidth}
\centering  
    \resizebox{0.7\textwidth}{!}{\includegraphics{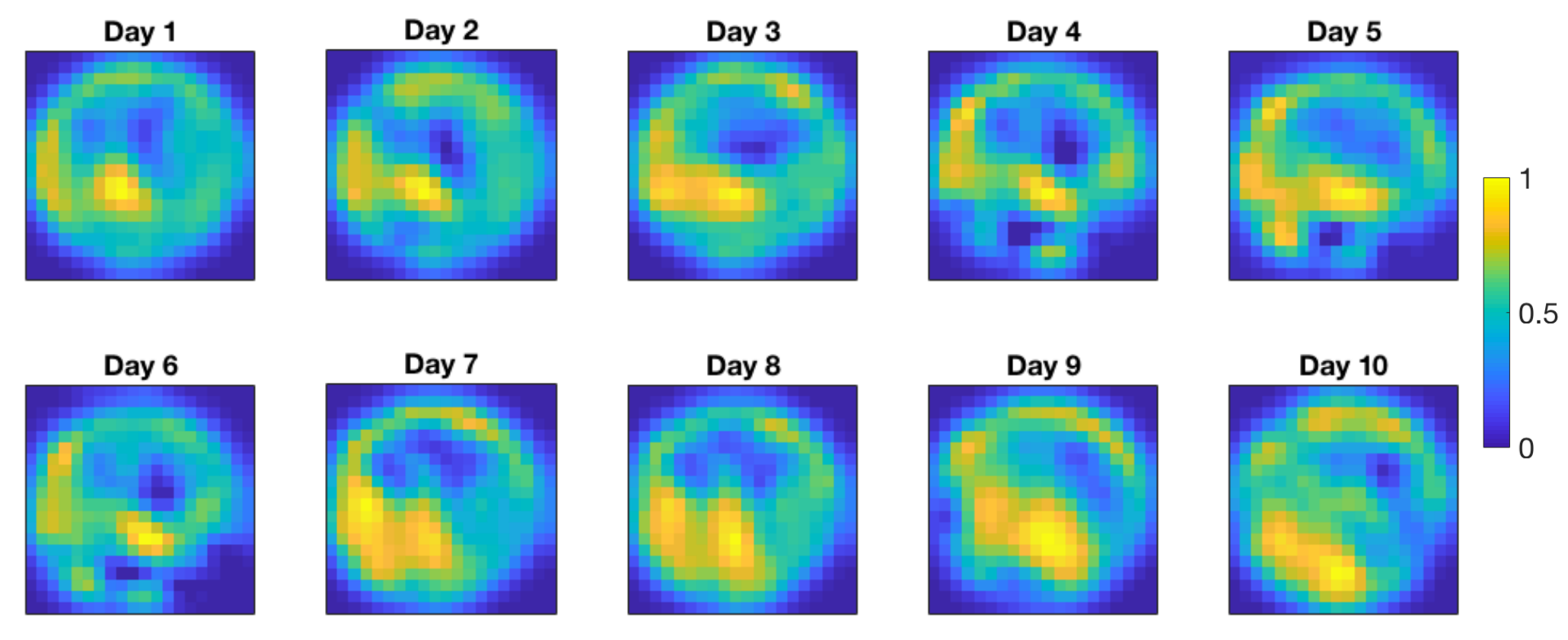}}
    \caption{Reconstruction of LCLS data by day. Reconstructions are relative to the center of the laser beam, so a shift in beam centroid on the cathode will correspond to a shift in the reconstructed cathode QE. There is a consistent hotspot just below and to the left of the laser centroid. The strong circular ring of high QE at the edge of the laser aperture may be an artifact from the reconstruction algorithm.}
    \label{fig:LCLS_recon_day}
\end{minipage}
\end{figure}

\noindent SNR of 71. We scan the hyperparameters to optimize the MSE of the reconstruction, and the results are presented in Fig.~\ref{fig:LCLS_sim_day}.

Using the same hyperparameters that optimize the reconstruction for the simulated QE, we feed the day-by-day data into the ADMM algorithm and obtain the reconstructed QE maps shown in Fig.~\ref{fig:LCLS_recon_day}. 
The results show a consistent hot spot at the lower part of the beam region. 
From the steering mirror movement shown in Fig.~\ref{fig:LCLS_M2}, between days 6 and 7, the laser moves left and up, which explains why the hotspot of the cathode QE moves right and down in the reconstructions over the ten days (Fig.~\ref{fig:LCLS_recon_day}). We note that a circular ring at the upper edge of the laser aperture appears in both simulation and reconstruction. Therefore, it may be an artifact of the reconstruction algorithm rather than a real feature.  

Without a dedicated measurement shift, we have no way to measure the ground truth QE for comparison with Fig.~\ref{fig:LCLS_recon_day}. However, by reconstructing a consistent QE map repeatedly over many days, we have some degree of cross-validation of the result. We also highlight that the reconstructed QE does not appear to be related to the\newline \newpage
\noindent average laser profile; the laser shape changes significantly during the data collection (Fig.~\ref{fig:LCLSlasers}) without changing the QE reconstructions in Fig.~\ref{fig:LCLS_recon_day}.

Finally, we emphasize that these reconstructions have used a relatively small number of measurements. Given the passive nature of the method, extensive data sets can be collected without interfering with operation. For example, at LCLS-II it will be feasible to measure millions of shots per days, enabling reconstructions with relatively small amounts of jitter in the laser.


\section{Conclusions}
In conclusion, we present a novel method for measuring the spatial features in QE of injector cathodes based on a ghost imaging reconstruction framework. The method is validated on ad-hoc measurements at the UCLA Pegasus photoinjector, showing equivalent results to a common raster scan. This experiment is used to qualitatively assess the variation in the illumination pattern needed to obtain a reliable reconstruction. The method is then applied to passive measurements taken during routine operation of LCLS to show the main advantage of being able to run parasitically off the normal operation. 

Cathode QE ghost imaging holds promise for electron-source based user facilities where dedicated time to study the performances of the injector is limited, and the spatial variation of the QE is a critical quantity affecting the final beam brightness and machine performance. More generally, we believe ghost imaging can be a valuable tool at accelerators, where noisy probes are common and opportunities for dedicated studies are limited \cite{lane2019advantages,ratner19cl}. The ghost imaging philosophy of ``measurement is easier than control'' can find applications for a wide range of accelerator problems.



\begin{acknowledgments}
We would like to thank Brendan O'Shea for inspiration to study ghost imaging for QE maps, and Lauren Alsberg, Franz-Josef Decker, Dave Dowell, and Tim Maxwell for helpful discussions. This work was partially supported by STROBE National Science Foundation Science \& Technology Center, Grant No. DMR-1548924.
\end{acknowledgments}

\appendix

\section{Simulation}
\label{app:simulation}

\begin{figure}[b!]
    \centering
    \includegraphics[width = 0.95\linewidth]{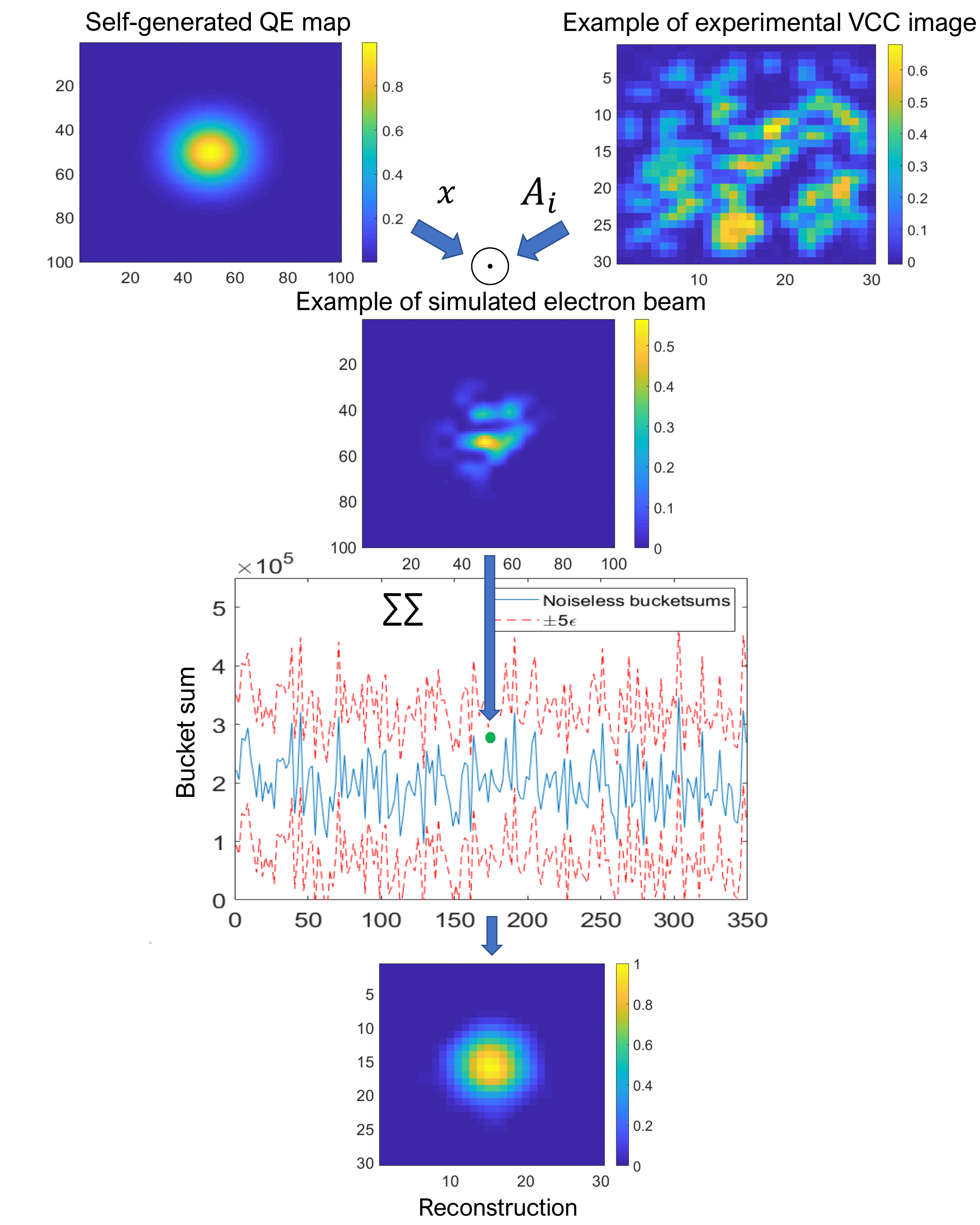}
    \caption{Overview of simulation using UCLA experimental data. An example of an experimental VCC image (top-right), a user-generated QE map (top-left) and simulated electron beam (top-middle) are shown. Bucket sums are calculated integrating all pixels of the simulated e-beam and and adding random Gaussian noise $\epsilon_i$, as described by Eq. \ref{eq:b=Ax_sim}. Note that the VCC image is upsized to match the size of the QE map. The bottom-middle plot shows example simulated bucket sums for all random scan measurements. The blue line shows the `noiseless' bucket sums and the red dashed line shows boundaries of $\pm 5 \epsilon$ within which the simulated bucket sums are contained. An example reconstruction using an experimental SNR of 7.8 and $N=$ 350 random masks is given at the bottom. The reconstruction is the same size as the original VCC images.}
    \label{fig:simulationFig}
\end{figure}

\begin{figure*}
\centering  
    \resizebox{0.8\textwidth}{!}{\includegraphics{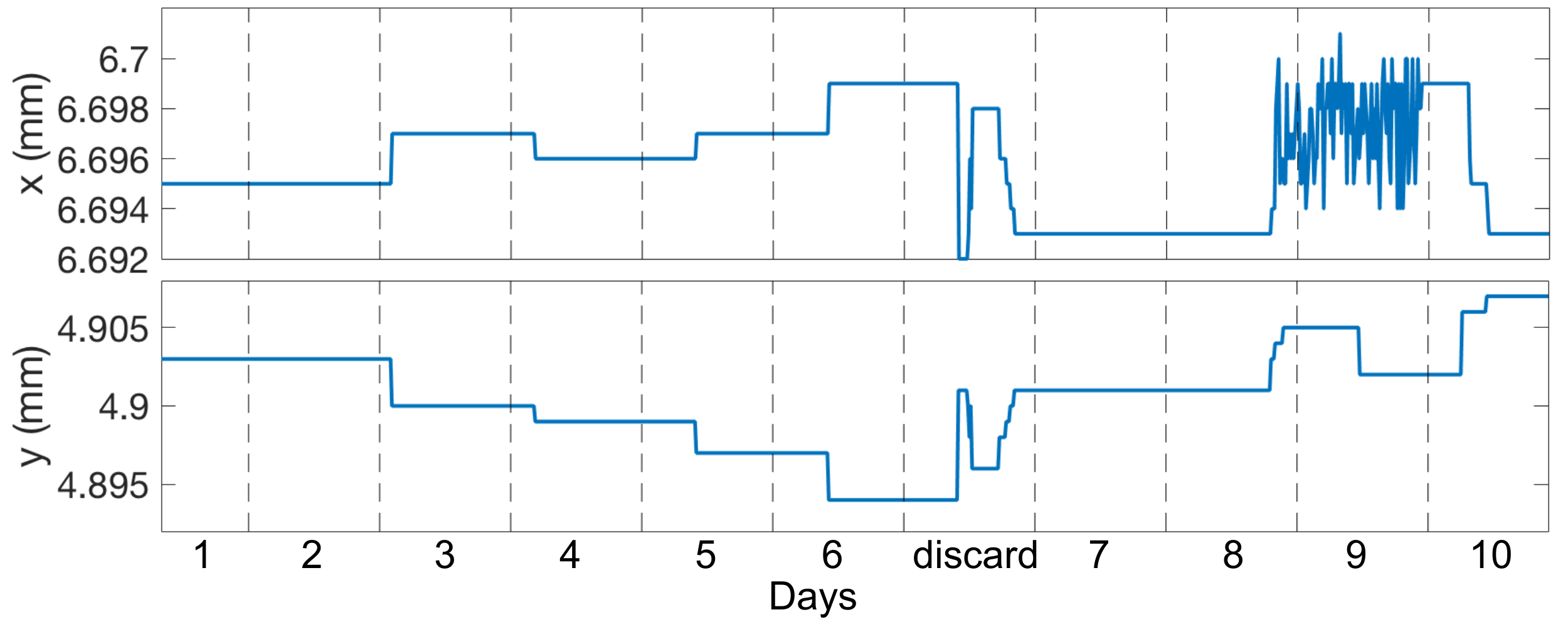}}
    \caption{Steering mirror movement across the days of data acquisition. The data between day 6 and 7 are discarded due to operational interruptions. Top panel: horizontal movement. Bottom panel: vertical movement. The positive sign in horizontal direction means the laser moves right, in vertical up.}
    \label{fig:LCLS_M2}
\end{figure*}

To select ADMM hyperparameters for the experimental reconstructions, as well as characterize the spatial variation of the measured laser profiles needed to resolve arbitrary QE maps, we conduct measurement simulations using the experimental VCC images and user-generated QE maps. An example simulation using UCLA data is described in Fig.~\ref{fig:simulationFig}. The bucket sum for the $i^{\mathrm{th}}$ example is calculated as 
\begin{equation}
    b_i = \mathbf{a}_i\mathbf{x} + \varepsilon_i
\label{eq:b=Ax_sim}
\end{equation}
where $\mathbf{a}_i$ is the $i^{\mathrm{th}}$ row of $\mathbf{A}$, and $\varepsilon_i = \mathcal{N}(0,\varepsilon)$ is zero-centered Gaussian noise with standard deviation $\varepsilon$.
The noise level, $\varepsilon$, is the bucket mean divided by the experimental SNR. As several bucket sums were obtained for a single mask during experiment, the experimental SNR was calculated as follows: 
\begin{equation}
    SNR = \frac{1}{N} \sum_{i=1}^{N} \frac{\mu_{i}}{\sigma_{i}}
    \label{eq:expSNR}    
\end{equation}
where $N$ is the total number of unique masks obtained during the experiment, and $\mu_i$ and $\sigma_i$ are the mean and standard deviation respectively of the bucket sums for a single mask. 

ADMM hyperparameters for reconstructions are selected by scanning through a range of values. Hyperparameters that correspond to the lowest MSE for simulated reconstructions are separately selected for different datasets. 

\section{LCLS mirror movement}

The movement of the steering mirror upstream of the VCC is recorded during the days of data acquisition (Fig.~\ref{fig:LCLS_M2}). The data between days 6 and 7 are discarded due to operational interruptions.


\nocite{*}
 
\bibliography{apssamp}

\end{document}